\documentclass{article}
\usepackage{spconf,amsmath,graphicx}
\usepackage{xcolor}
\usepackage{amsmath,bbm}
\usepackage{amssymb}
\usepackage{booktabs}
\usepackage{multirow}
\usepackage{cite}
\usepackage[mathscr]{euscript}
\usepackage{fancyhdr}

\DeclareMathAlphabet\mathbfcal{OMS}{cmsy}{b}{n}

\usepackage{fancyhdr}
\fancypagestyle{firstpage}{
  \fancyhf{}
  \fancyhead[C]{\vspace{-2.2cm} \small Accepted for presentation at the IEEE Int. Conf. on Acoustics, Speech and Signal Processing (ICASSP), Toronto, Canada, June 6-11, 2021.\\© 2021 IEEE. Personal use of this material is permitted. Permission from IEEE must be obtained for all other uses, in any current or future media, including reprinting/republishing this material for advertising or promotional purposes, creating new collective works, for resale or redistribution to servers or lists, or reuse of any copyrighted component of this work in other works. \vspace{0.3cm}} 
}

\title{Collaborative Intelligence: Challenges and Opportunities}
\name{Ivan V. Baji\'{c},$^\dagger$ \hspace{1pt} Weisi Lin,$^\ddagger$ \hspace{1pt} and Yonghong Tian$^\mathsection$}
\address{$\dagger$ School of Engineering Science, Simon Fraser University, Canada\\
$\ddagger$ School of Computer Science and Engineering, Nanyang Technological University, Singapore\\
$\mathsection$ Department of Computer Science and Technology, Peking University, China 
}

\begin{document}

\maketitle


%
\begin{abstract}
This paper presents an overview of the emerging area of collaborative intelligence (CI). Our goal is to raise awareness in the signal processing community of the challenges and opportunities in this area of growing importance, where key developments are expected to come from signal processing and related disciplines. The paper surveys the current state of the art in CI, with special emphasis on signal processing-related challenges in feature compression, error resilience, privacy, and system-level design.

\end{abstract}

\thispagestyle{firstpage}

\section{Introduction}
\begin{figure}
    \centering
    \includegraphics[width=\columnwidth]{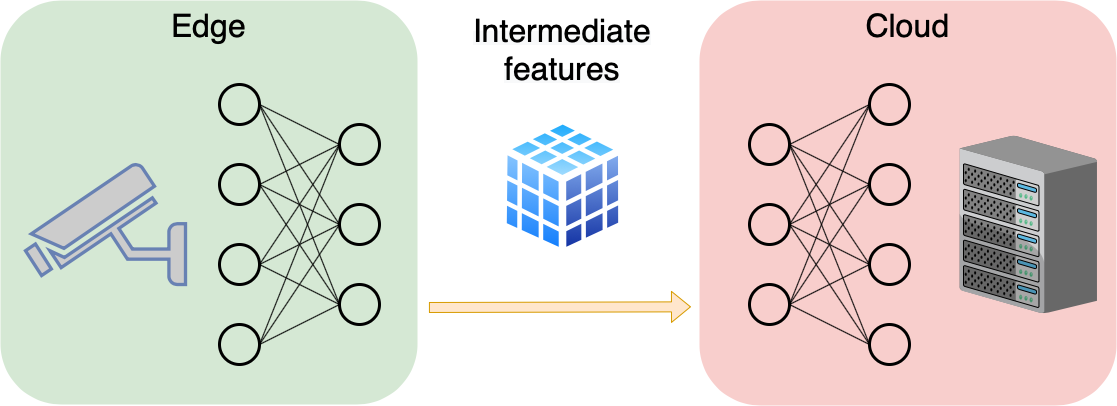}
    \caption{Basic collaborative intelligence system}
    \label{fig:CI_system}
\end{figure}

Artificial Intelligence (AI) is moving from research labs to the real world. One of the promising avenues for bringing AI ``to the edge'' is \emph{Collaborative Intelligence} (CI), a framework in which AI inference is shared between the edge devices and the cloud. In CI, typically, the front-end of an AI model is deployed on an edge device, where it performs initial processing and feature computation. These intermediate features are then sent to the cloud, where the back-end of the AI model completes the inference, as shown in Fig.~\ref{fig:CI_system}. Variations on this basic model are possible, where multiple edge devices or multiple inference engines are involved.

CI has been shown to have the potential for energy and latency savings compared to the more typical cloud-based or fully edge-based AI model deployment~\cite{kang2017neurosurgeon, jointdnn}, but it also introduces new challenges, which require new science and engineering principles to be developed in order to achieve optimal designs. In CI, a capacity-limited channel is inserted in the information pathway of an AI model. This necessitates compression of features computed at the edge sub-model. Moreover, errors introduced into features due to channel imperfections would need to be handled at the cloud side in order to perform successful inference. Finally, issues related to the privacy of transmitted data need to be addressed. 

This paper presents an overview of signal processing-related challenges an opportunities brought by CI. Section~\ref{sec:feature_compression} presents an overview of the existing work and future challenges in deep intermediate feature compression. Section~\ref{sec:error_resilience} discusses error resilience of intermediate features to bit errors and packet loss, while Section~\ref{sec:privacy} focuses on privacy aspects. Section~\ref{sec:system_design} discusses challenges in system-level design of CI systems. Finally, Section~\ref{sec:conclusions} concludes the paper.

\section{Intermediate feature compression}
\label{sec:feature_compression}
Due to the capacity-limited channel in the information path of a CI model, intermediate features that are being transferred from the edge to the cloud need to be compressed. Feature compression in itself is not a new topic; earlier work on Compact Descriptors for Visual Search (CDVS)~\cite{CDVS_TIP_2016} and Compact Descriptors for Video Analysis (CDVA)~\cite{CDVA_MM_2019} have considered compression of (mostly handcrafted) features for visual analytics, especially matching, search, and retrieval. Such features are  ultimate ones, because they can be directly used in the cloud (without the need of significant back-end processing as in Fig. 1).  On the other hand, in CI, the focus is on learned, intermediate features, and the applications range beyond visual analytics.  

A new scheme of coding and transmitting intermediate deep learnt features instead of sensor data or the aforementioned ultimate features is emerging recently~\cite{Chen19,dfc_for_collab_object_detection, Choi2018NearLosslessDF}, where deep neural networks are split into the edge part and the cloud one respectively. A task is performed in an edge-cloud collaborative manner. The generic backbone networks with most computing burden in visual analysis may be located on the edge to extract intermediate features, while the task-specific components are then assigned to the cloud. The load balancing can be achieved in a flexible and task-specific manner.  A generic deep learning architecture can be deployed to reduce the cost of edge devices.
To enable the new paradigm, compression and coding techniques for intermediate deep learning features are 
needed.

Lossless compression is not practical for intermediate deep feature coding~\cite{Chen19b}, and most existing works have employed conventional video codecs to perform lossy feature compression. Research in~\cite{dfc_for_collab_object_detection, Choi2018NearLosslessDF} has performed lossy intermediate deep feature compression for the object detection task, utilizing traditional image/video codecs including PNG, JPEG, JPEG2000, VP9 and HEVC.
A coding framework with three modules (PreQuantization, Repack, and VideoEncoder) has been devised \cite{Chen19,Chen19b} to encode intermediate deep features and demonstrated the compression performance on three computer vision tasks (image classification, image retrieval, and visual object detection). In addition, there is investigation  adopting multi-task learning~\cite{Alvar19}.

The optimization goal of video coding is to minimize signal loss under certain bitrate constraints. In contrast, feature coding's optimization goal is to minimize the loss of analysis and recognition at a given bit rate. Up to date, solutions toward  deep video feature compression, with the analysis or retrieval performance being maintained, still remains open. A rate-performance-loss optimization model~\cite{Ding2017RatePerformanceLossOF} is to make a tradeoff between required bitrate and analysis/retrieval performance. It is similar to the Rate-Distortion Optimization (RDO) in video coding but with the performance loss as the distortion term. To take advantages of temporal redundancy in deep features for improved compression performance, three types of features are defined according to  potential coding modes: Independently-coded feature (I-feature); Predictively-coded feature (P-feature); and Skip-coded feature (S-feature). Following this, a joint coding framework for local and global deep features extracted from videos was proposed, in which local feature coding can be accelerated by making use of the frame types determined with global features, while the decoded local features can be used as a reference for global feature coding~\cite{Ding2020JointCO}. The authors in~\cite{Wang2020TowardsAF} introduced feature and texture compression in a scalable coding framework, where the base layer is catered for the deep learning feature while the enhancement layer aims to reconstruct the texture. Towards collaborative compression and intelligent analytics, Video Coding for Machines (VCM)~\cite{MPEG_AHG_VCM,Duan2020VideoCF} bridges the gap between feature coding for machine vision and video coding for human vision.

The feature (instead of the whole visual signal) compression enables more accurate visual features since they are extracted from original  (rather than decoded) visual signal. As noted earlier, there is flexibility of computational load balancing between edge devices and cloud servers. There are fewer privacy concerns because not the whole video is coded and transmitted, as discussed further in Section~\ref{sec:privacy}. There has been initial standardization effort for intermediate feature compression, as noted above: MPEG has started an ad-hoc working group on VCM, and AVS (Audio Video Coding Standard of China) has had a visual feature coding working group for intermediate deep learnt feature compression earlier~\cite{Ma17}. 

The coding framework described above adopts traditional video codecs, which do not closely match the feature data distribution since they have been designed and optimized for natural images and videos; and some  coding tools are inevitably redundant in feature compression, consuming unnecessary computing resource. For the next steps,  more coding tools designed for intermediate deep features are anticipated. 

Traditional coding methods aim for best video quality under certain bit-rate constraint for human consumption while the said new framework allows to extract useful information directly for machine analysis. Hence, joint optimization of video coding and feature coding can be investigated in the future. We can explore how to use both feature bitstream and video bitstream to reconstruct high quality videos for human or/and machine intelligence.

As to the deep learning models, coding-aware backbone networks and task-specific model design is another meaningful research direction. In particular, coding-aware backbone networks at the edge side should be able to generate features that are robust against lossy compression. The coding-aware task-specific model at the cloud side should maintain high performance with possibly-corrupted features. For instance, coding-aware feature generation has been achieved 
in~\cite{Eshratifar19}.

\section{Error and loss resilience}
\label{sec:error_resilience}
Besides being capacity-limited, the communication channel will introduce errors into the information path of a CI model. At the physical layer, bit errors will be observed in the transmitted data. At the network/transport layer, packet loss will be introduced  into the transmitted data. Error control methods will need to be designed to handle these impairments. 

From the existing work on feature compression, it appears that intermediate features can tolerate some degree of quantization noise without much degradation of the model's accuracy. However, resilience of intermediate features to bit errors and packet loss, and the corresponding error control methods, are currently largely unexplored. Among the few existing works,~\cite{choi_neural_2019, BottleNet++} studied joint source-channel coding of intermediate features,~\cite{DFTS_2018} explored simple interpolation methods for recovering features missing due to packet loss, while~\cite{Bragile_2020} proposed low-rank tensor completion for this purpose. All these methods have considered single-image input to the CI system, rather than video input. There are many open issues here, revolving around finding the best ways to exploit various redundancies (spatial, temporal, inter-channel) in the feature domain, to come up with effective data recovery methods. 

One could also imagine developing unequal error protection schemes, where the level of protection would be tailored to the importance of specific features in relation to the model accuracy. Gauging the feature importance may be achieved using attention mechanisms~\cite{zhang_residual_2018}, while the actual error protection may be implemented via channel codes of varying strength, unequal power allocation, or by matching features of different importance with appropriate network traffic management policies and Quality of Service (QoS) schemes. 

\section{Privacy}
\label{sec:privacy}
From the privacy point of view, an inherent advantage of CI systems over more traditional cloud-based analytics is that the original input signal never leaves the edge device; only intermediate features are transmitted for further processing. These intermediate features may or may not resemble input images, but even if they look very different from input images, this does not mean that it is impossible to reconstruct some private information from them. Privacy is an extensively studied topic in machine/deep learning~\cite{mireshghallah_privacy_2020}. However, much of the work in this area is focused on learning, whereas CI approach requires inference-time privacy. Based on the classification presented in~\cite{mireshghallah_privacy_2020}, CI privacy would be most related to \emph{model inversion} and \emph{attribute inference} type of attacks.  

\begin{figure}
    \centering
    \includegraphics[width=\columnwidth]{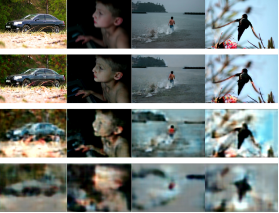}
    \caption{Adapted from~\cite{Choi2018NearLosslessDF}. Input images (top row) and reconstructions from 7$^\text{th}$, 11$^\text{th}$, and 17$^\text{th}$ layer of YOLOv2~\cite{YOLO2}.}
    \label{fig:input_reconstruction}
\end{figure}

Fig.~\ref{fig:input_reconstruction}, adapted from~\cite{Choi2018NearLosslessDF}, shows input images reconstructed from 7$^\text{th}$, 11$^\text{th}$, and 17$^\text{th}$ layer of YOLOv2~\cite{YOLO2}. Input reconstruction was accomplished using a trained ``mirror model,'' which has the same architecture as the front-end model up to the given layer, but with pooling layers replaced by upsampling, and convolutions replaced by transpose convolutions. Although this is not necessarily the best architecture for input reconstruction, it does show that fairly decent reconstruction, potentially revealing private information, is possible from the 7$^\text{th}$ layer features of YOLOv2, and perhaps 11$^\text{th}$ layer as well in some cases. Further, it should be noted that full input reconstruction may not be necessary for privacy to be breached; for example, someone's identity may be revealed via the shape of heir head or the silhouette of their haircut, even without the fine details of their facial features. 

Hence, the privacy problem in CI systems is much more involved than simply avoiding transmission of the input signal. However, since the intermediate features are learnt, this creates opportunities to construct a learning process that would result in features that are privacy-friendly in addition to being able to support the main analytics task(s). One way to accomplish this would be to develop loss functions that would increase the uncertainty (entropy) about private information while reducing (or trading off) the error(s) on the main task(s). Another approach would be to create noise to be added to the intermediate features to improve their privacy~\cite{mireshghallah_shredder_2020}. One could also imagine an adversarial learning strategy~\cite{GAN_NIPS2014}, where the ``discriminator'' would try to infer private information from intermediate features, while the ``generator'' would try to create features that protect it.

\section{System-level design}
\label{sec:system_design}
Compared with the traditional cloud-centered data processing mode, the CI system faces many challenges for big data processing, such as latency constraints, and memory occupation. Big data processing for CI relies on huge computing resources. The traditional cloud-centered processing architecture brings an enormous burden to the cloud, especially with the high computational complexity of deep neural networks (DNNs). Meanwhile, AI-driven devices are increasingly expected to perform various related tasks by running multiple homogeneous DNNs simultaneously. It is a big challenge to deploy these networks on resource-constrained devices.

Typically, a DNN can be divided into several modules. In this way, part of the computation of DNN modules can be migrated to the front-end devices that are equipped with graphic processing units (GPUs), which can alleviate the cloud burden and shorten the processing latency. A DNN partition method was introduced by Kang \emph{et al.}~\cite{kang2017neurosurgeon} to improve the efficiency and throughput of the whole system. A scheduler was developed to partition the DNN computation between mobile devices and data centers automatically. Eshratifar~\emph{et al.}~\cite{Eshratifar19} presented an intermediate feature compression method for the dynamic DNN partition. An inference computation allocation method among multi-layer Internet of things systems was introduced in \cite{Zhou2019AAIoTAA}, in which all the videos still need to be sent to the cloud servers. Zhang \emph{et al.}~\cite{Zhang2018CooperativeCompetitiveTA} provided a task-level allocation method to shorten the end-to-end latency.

For the multi-task memory storage reduction, sharing information on different networks provides a promising way to reduce the sizes of multiple correlated models. Therefore, how to optimize the storage usage of multi-task homogeneous networks without significant performance degradation in each task is a tricky problem. It is necessary to share parts of different networks to remove inter-redundancy information. Most studies focus on compressing a single deep network~\cite{zhang2016accelerating,he2019filter,zhuang2018towards,zhang2018lq}, although removing the intra-redundancy within every single model is important for their usage on the resource-constrained devices.  An intuitive solution to reduce the inter-redundancy is multi-task learning techniques~\cite{misra2016cross,kendall2018multi}. Multi-task learning aims to improve generalization by sharing representations between related tasks. A commonly-used multi-task learning approach is to share the hidden layers between all tasks while preserving several task-specific output layers, or even use a single shared architecture ~\cite{georgiev2017low, HashimotoA, KaiserOne}. Recently, more works pay attention to develop more effective mechanisms for multi-task learning by learning additional task-specific parameters for new tasks. Additional cross-stitch units was introduced in ~\cite{misra2016cross} to allow features to be shared across tasks. In~\cite{RebuffiLearning}, a residual adapter module was proposed to enable a high degree of parameter sharing between domains.The authors in ~\cite{Incremental} learned new filters that are linear combinations of existing filters for new tasks. A Multi-Task Attention Network consisting of two major components was proposed in ~\cite{Liu2019}: a single shared network learning a global feature pool and a soft-attention module for each task, allowing for automatic learning of both task-shared and task-specific features. 

Most of the multi-task learning methods adopt pre-defined shared structures~\cite{caruana1993multitask,long2015learning,georgiev2017low}, which may result in performance degradation due to improper sharing of knowledge. Some approaches learn what to share across tasks from pre-trained models~\cite{he2018multi, ChouUnifying,unknown}. However, such methods suffer from some limitations: 1) \textit{The dependence on pre-trained models} makes these methods lacking in versatility and adaptability to different initialization of models. 2) \textit{Additional training overhead} is required to obtain the pre-trained models when some pre-trained models are not available.

Meanwhile, some other popular methods~\cite{he2018multi, ChouUnifying, unknown} have been explored to share parameters among pre-trained models. Multiple well-trained DNNs are merged in ~\cite{he2018multi} and~\cite{ChouUnifying} via weight sharing directly. A branched multi-task structure was proposed~\cite{unknown} by leveraging the employed task affinities for layer sharing among pre-trained networks. However, these methods are over-reliant on the pre-trained models, resulting in a lack of adaptability to different initialization models and additional training overhead.

\section{Conclusion}
\label{sec:conclusions}
In this paper we presented a brief overview of signal pro-cessing-related challenges and opportunities in collaborative intelligence. These were grouped into areas where we felt the signal processing community will make key contributions: feature compression, error resilience, privacy, and system-level design. Many other important areas, such as lightweight/reduced-precision deep models, embedded DNN system development, feature communication and scheduling, could not be included due to space constraints, but in those areas too, signal processing is expected to provide key insights and solutions. We look forward to seeing the field grow, and signal processing playing an important role in it. 
\section{Acknowledgement}
Prof. Tian's work is partially supported by grants from the National Natural Science Foundation of China under contract No. 61825101.

\small
\bibliographystyle{IEEEbib}
\bibliography{ref}

\end{document}